# Technical manual: a survey of scintillating medium for high-energy particle detection

Adil Baitenov [1], Dmitriy Beznosko, Alexander Iakovlev

*Department of Physics, Nazarbayev University, Astana, Kazakhstan*
[1] *abaitenov@nu.edu.kz*

*Abstract*—There are various particle detection methods used nowadays and the most common is using scintillators. Among scintillating materials, solid plastic and water-based liquid scintillators (WbLS) are the latest development. In particular, WbLS allows researchers to apply different particle detection methods for increased experiment efficiency. This survey attempts to make an overview on detection methods and detectors in high-energy physics using scintillators. It is meant as a summary for those new to scintillator detectors and looking for general material on the topic.

## I. INTRODUCTION

High-energy physics studies fundamental particles and radiation. Despite of the first suggestions about existence of elementary particles from the $6^{th}$ century BC, actual particle physics history starts from John Dalton and his work on stoichiometry that concluded that every element consists of unique atom in $19^{th}$ century. In order to learn the nature of the elementary particles, scientists needed detection techniques that could help them with observation. Detection methods can be divided into observation-based old and electronic readout-based new generation methods. Old generation methods include: Bubble chamber (uses overheated liquid), Cloud chamber (overcooled vapor), Photographic plates (AgI crystals), Spark chamber (air); and new generation methods are: Gaseous ionization detectors (compressed gas), Cherenkov detectors and Scintillation detectors. Among the modern methods, most common detection technique is scintillation-based detectors. There are different types of scintillators that can be divided into two main groups: Inorganic and Organic. This paper focuses on organic scintillators, solid and liquid ones, including water-based since they are the latest development.

## II. ORGANIC SCINTILLATORS

### A. Scintillators explained

First, we need to understand what scintillator is and how it works. When a charged particle passes through any medium, it deposits some amount of its kinetic energy in that material as the energy of the excited electrons via electromagnetic interaction with them. Scintillator is the type of material, in which the electrons de-excite mostly with photon emission and not thermally with phonon excitation. Sometimes a metastable state is encountered, which means that relaxation from the excited state is delayed. That leads to one or both of the following phenomena: fluorescence and phosphorescence that together comprise the scintillation process.

### B. Particle detectors using organic scintillators

The particle detectors using the scintillating materials are often referred to as counters as their output is proportional to number of particles in the simplest case. Note that in many cases they don't only count the number of passing particles but can infer their species and direction and speed of travel if more than one counter is used. Addition of external magnetic field makes the measurement of the particle momentum possible as well.

Scintillation counter consists of scintillator material and a light sensor that detects photons caused by passing of charged particle, and converts them into electric pulses that are then amplified and analyzed to get information about the detected particle. Some typical photon detectors are PMT (photo multiplier tube), or more recent Geiger mode silicon sensors [1] [2] [3]. The peak detection efficiency for these sensors is typically in the blue to green parts of the visible spectrum.

However, typical organic scintillator emits photons in UV range, and, in order for more efficient detection, it must be shifted to visible spectrum. For that purpose, one or more wavelength shifter is added to scintillator. Wavelength shifter is photofluorescent material that absorbs higher energy photons and emits lower energy photons. Typically, the range of the shift is narrow, therefore two consecutive shifters are used as a rule.

### C. Types of organic scintillators

Organic scintillators are based on hydrocarbon compounds that contain benzene ring structures. Typically, de-excitation of organic scintillators is fast and occurs on a ~10 ns scale, which is sufficient for most detection purposes. Examples of liquid organic scintillators are PC (pseudocumene or 1.2.4-trimetilbenzene) and toluene ($C_7H_8$).

In organic scintillators, scintillation light is a product of transitions by the free valence electrons occupying molecular orbitals (typ. π-electron of the benzene ring). Charged particle going through the scintillator excites the lowest triplet state $T_0$ to $T^*$, which decays without photon emission. Then $T_0$ decays to excited level of singlet state $S^*$ by interacting with another $T_0$ molecule. $S^*$ decays to $S_0$ emitting photons in UV range (see Figure 1) [4].

Plastic scintillators consist of scintillator base and normally two wavelength shifters (called fluor and shifter) that are suspended in the base solid polymer matrix. Commonly used fluor is PPO (2.5-diphenyloxazole) and the shifter is POPOP (1.4-di-(5.phenyl-2-oxazolyl)-benzene).

Liquid scintillator (LS) consists of organic solvent base in which same fluor and shifter are dissolved. Examples of liquid scintillators are pseudocumene (1.2.4-trimetilbenzene) and LAB ($C_6H_5CHC_{12}H_{25}C_{15}H_{31}$).

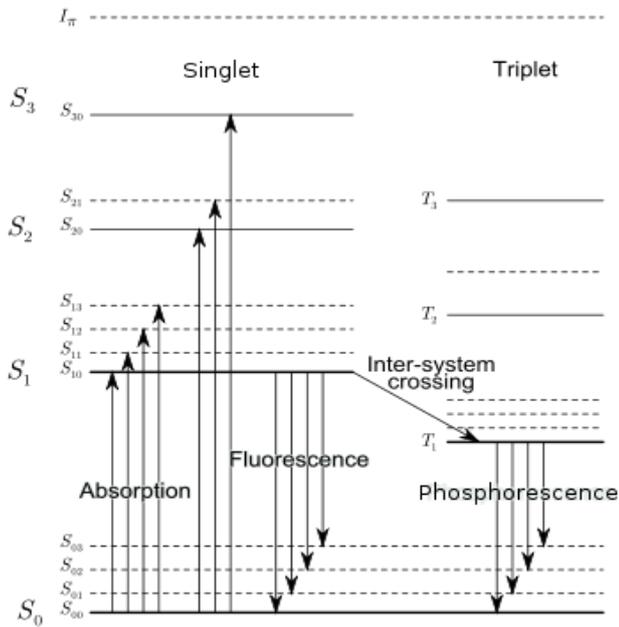

**Figure 1: Scintillation in organic materials [4]**

### III. SOLID AND WATER-BASED LIQUID SCINTILLATORS

One of the main reasons that liquid scintillators may be preferable over solid scintillators is that in order to have efficient number of detections and properties measurements of particles with low interaction cross-section (such as neutrinos), such detector should contain a large continuous volume of active material, e.g. scintillator. Solid scintillators are used only in case if each active particle detector can be small-scaled or modular [5] [1], and can also include the wavelength shifting fiber [6] for better light collection (a sample schematic is shown in Figure 2).

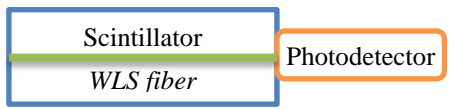

**Figure 2: Solid scintillator module typical design schematic.**

In addition, the response of the organic scintillator is unaffected by the strong magnetic field [7] [8]. The particle track of course is affected, changing the amount of the energy deposited. This allows using such modules to build even very large detectors systems, such as T2K Pi-zero detector [9].

However, if researchers are using a detector similar to Super-Kamiokande (SK) [10], a 50,000 ton volume detector located in Japan, liquid scintillator is the only option as it can fill the volume of detector chamber. Note that there are other problems that scientists encountered. First, in order to implement particle identification and the reconstruction of its direction and energy, Cherenkov radiation produced by the particle must be detected in SK. However, all charged particles below Cherenkov threshold are missed [11] (for a given particle, Cherenkov threshold is the energy at which particle moves faster than the speed of light in the current medium). The second problem is that liquid scintillator absorbs the Cherenkov light (which is mostly in UV region) and re-emits it isotropically, thus destroying the information about the particle via this channel. In addition, filling SK detector will cost estimated $50-70M; there are also safety issues since organic scintillators consist of aromatic solvents, which are flammable or even combustible (Figure 3) [12]. Because of a low freezing point, LS maybe considered for the Horizon-T cosmic rays detector system [13] [14] or similar [15].

*A. Water-based Liquid Scintillators*

In order to address problems mentioned above, new scintillator medium has been developed. WbLS is made by dissolving a certain percentage of LS in water using a surfactant agent. Advantages of WbLS in comparison with ordinary liquid scintillator are: preserved Cherenkov light cone and ability to detector the particles below the Cherenkov threshold via scintillator. Also, due to mixing of scintillator with water, WbLS is not flammable and its cost is scaled down since less scintillator itself is used. The rough estimate of filling Super-Kamiokande detector with WbLS is in the range of ten present or so of the pure scintillator cost. The current samples of WbLS appear to be stable in time [16].

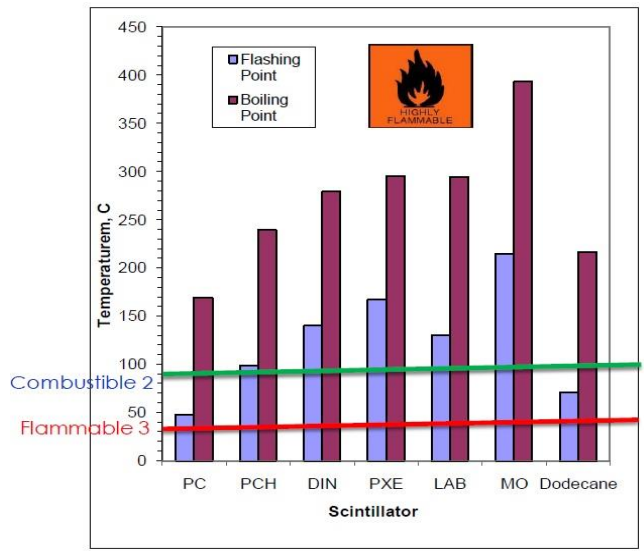

**Figure 3: Boiling and Flash points of liquid scintillators [12]**

## B. Current Disadvantages

- In SK or similar detector, super-pure water is required to keep attenuation of Cherenkov light as low as possible. If WbLS is used instead, it is not yet clear how to purify it over time.
- Due to lower percentage of pure scintillator used, the amount of photons produced is lower.
- Not applicable or hard to use in small scale experiments due to low scintillator light yield and inability to use Cherenkov light information effectively as energy resolution scales with coverage.

## IV. CURRENT AND FUTURE RESEARCH

By now, there is a number of research articles on WbLS is published. In 2008, Physics department of Carleton University, Canada, published their research on wavelength shifters (WLS) for Water Cherenkov detectors, in which they studied the way to significantly improve the light yield of a water-based Cherenkov detector by adding different wavelength shifters [17]. Authors tested several wavelength shifter compounds for possible deployment in the Sudbury Neutrino Observatory (SNO). Test results on optical properties and chemical compatibility for a few WLS candidates were reported. As a result of this research, a small-scale (few meters) Cherenkov detector was built to investigate the optical properties of WLS further [17].

In 2014, a large collaboration of researches published a concept paper: "Advanced Scintillator Detector Concept (ASDC): A Concept Paper on the Physics Potential of Water-Based Liquid Scintillator" [18], in which they proposed a concept of a new kind of large-scale detector capable of a very broad program of physics base on the developed WbLS and high-efficiency and high-precision-timing light sensors.

In late 2015, researchers of Brookhaven National Laboratory (BNL), USA, published an article on characterization and modeling of a Water-based Liquid Scintillator [16], in which they characterized WbLS using low energy protons, and also developed and validated a simulation model that describes the behavior of WbLS in detector for proton beam energies of 210 MeV, 475 MeV, and 2 GeV and for two WbLS compositions with water and pure LS used as reference. Results allowed estimation of light yield and ionization quenching of WbLS, as well as to understand the influence of the wavelength shifting of Cherenkov light on their measurements. These results are relevant to the suitability of WbLS materials in SK for the proton decay detection [16] and for next generation intensity frontier experiments.

Same BNL group conducted their research on measurement of radiation damage of WbLS and LS [19]. In this article, authors studied what effect of radiation damage to the scintillator will have upon application. They performed measurements of the degradation of light yield and optical attenuation length of LS and WbLS after irradiation by 201 MeV proton beams that deposited doses of approximately 52 Gy, 300 Gy, and 800 Gy in the scintillator. LS and WbLS exhibited light yield reductions of 1.74 +/- 0.55% and 1.31 +/- 0.59% after ≈ 800 Gy of proton dose, respectively. Result of this research have shown that scintillators would exhibit a systematic light yield reduction of approximately 0.1% after a year of operation [19].

In December of 2015, researchers of Tsinghua University, China, published their research on separation of scintillation and Cherenkov lights in WbLS based on Linear Alkyl Benzene, in which presents their observations on a separation of scintillation and Cherenkov lights in linear alkyl benzene sample. Separation of scintillation and Cherenkov lights is important feature for future neutrino and proton decay researches [20].

## A. Future research

Future research plans of scientists include:

- Develop an effective purification of WbLS on large scales.
- Optimize the composition of WbLS for more efficient charged particles detection and higher sensitivity.
- Check the stability of WbLS for longer periods of time.
- Proof WbLS usability in large scale detectors like Super-Kamiokande.

## V. CONCLUSION

Scintillators are materials that allow detector of a moving charged particle by emitting light that is proportional to the number of particles and their properties, such as the total energy. Organic scintillators emit light in a fast manner (ns time scale) and can be used for detection of fast processes. Organic scintillator can be solid or liquid, thus allowing for modular solid detector design, or large volume detector design.